\documentclass[PRL,twocolumn,showpacs,superscriptaddress,amsmath,amssymb]{revtex4}

\usepackage{graphicx}
\usepackage{dcolumn}
\usepackage{bm}
\usepackage{epstopdf}

\def\Mpl{M_{P}}

\newcommand{\beq}{\begin{equation}}
\newcommand{\eeq}{\end{equation}}
\newcommand{\beqa}{\begin{eqnarray}}
\newcommand{\eeqa}{\end{eqnarray}}

\newcommand{\bk}{{\mathbf{k}}}
\newcommand{\bx}{{\mathbf{x}}}

\newcommand{\nn}{\nonumber}

\def\TT{{\rm TT}}

\newcommand{\mn}{{\mu\nu}}

\def\de{\delta}

\def\la{\lambda}

\def\si{\sigma}

\def\De{\Delta}

\def\La{\Lambda}

\newcommand{\ben}{\begin{equation}}
\newcommand{\een}{\end{equation}}
\newcommand{\bea}{\begin{eqnarray}}
\newcommand{\eea}{\end{eqnarray}}
\newcommand{\ba}{\begin{array}}
\newcommand{\ea}{\end{array}}
\newcommand{\bit}{\begin{itemize}}
\newcommand{\eit}{\end{itemize}}
\newcommand{\vev}[1]{\left\langle#1\right\rangle}

\newcommand{\gw}{GW}
\newcommand{\Gws}{Gravitational waves}
\newcommand{\gws}{GWs}

\newcommand{\rhogw}{\rho_{\text{GW}}}
\newcommand{\Omgw}{\Omega_{\text{GW}}}

\newcommand{\uetcT}{C^T}
\newcommand{\etcT}{E^T}
\newcommand{\integral}{F^T}

\newcommand{\tRef}{t_{\rm ref}}
\newcommand{\tEnd}{t_{\rm end}}

\begin{document}

\title{Exact Scale-Invariant Background of Gravitational Waves from Cosmic Defects}

\author{Daniel G. Figueroa}
\affiliation{D\'epartement de Physique Th\'eorique and Center for Astroparticle Physics, Universit\'e de Gen\`eve, 24 quai Ernest Ansermet, CH--1211 Gen\`eve 4, Switzerland}
\affiliation{Physics Department, University of Helsinki and Helsinki Institute of Physics, P.O. Box 64, FI-00014, Helsinki, Finland}
\author{Mark Hindmarsh}
\affiliation{Physics Department, University of Helsinki and Helsinki Institute of Physics, P.O. Box 64, FI-00014, Helsinki, Finland}
\affiliation{Department of Physics and Astronomy, University of Sussex, Brighton BN1 9QH, U.K.}
\author{Jon Urrestilla}
\affiliation{Department of Theoretical Physics, University of the Basque Country UPV/EHU, 48080 Bilbao, Spain}
\affiliation{Department of Physics and Astronomy, University of Sussex, Brighton BN1 9QH, U.K.}

\date{\today}

\begin{abstract}
We demonstrate that any scaling source in the radiation era produces a background of gravitational waves with an exact scale-invariant 
power spectrum. Cosmic defects, created after a phase transition in the early Universe, are such a scaling source. We emphasise that the result is independent of the topology of the cosmic defects, the order of  phase transition, and the nature of the symmetry broken, global or gauged. As an example, using large-scale numerical simulations, we calculate the scale invariant gravitational wave power spectrum generated by the dynamics of a global O($N$) scalar theory. The result approaches the large $N$ theoretical prediction as $N^{-2}$, albeit with a large coefficient. The signal from global cosmic strings is $\mathcal{O}(100)$ times larger than the large $N$ prediction.
\end{abstract}


\maketitle

Information about new physics in the very early universe is directly accessible through weakly-interacting relics. The weakest of all interactions is gravity, and sufficiently energetic processes leave behind characteristic signatures in relic gravitational waves (GW). New {\gw} observatories are under construction 
or have been proposed 
 \cite{MaggioreBook,Sathyaprakash:2012jk}. These are very likely to detect astrophysical sources soon, and have the potential to detect cosmological sources or strongly constrain early universe scenarios.

{\Gws} are produced whenever there is an energy-momentum tensor with a transverse-traceless (\TT) part~\cite{MaggioreBook}. In the absence of any source, GW are also generated quantum mechanically during inflation~\cite{Starobinsky:1979ty}. In this paper we study the GW background 
generated by a scaling source, whose energy-momentum tensor is 
proportional to the square of the Hubble parameter. 

Scaling is exhibited by  cosmic defects, products of  a phase transition in the early universe~\cite{Kibble:1976sj}.  If the vacuum manifold $\mathcal{M}$ is topologically non-trivial, i.e.~has a non-trivial homotopy group $\pi_n(\mathcal{M}) \neq \mathcal{I}$, topological field configurations  arise: strings for $n\!=\!1$, monopoles for $n\!=\!2$, and textures for $n\!=\!3$. For higher $n$, non-topological field configurations are created. When the spontaneously broken symmetry is global, the defects generated are termed global. When the broken symmetry is gauged, local defects appear. Global defects (independently of their topology), as well as cosmic strings (global or gauged) exhibit scaling behaviour, sufficiently long after the phase transition that created them~\cite{Turok:1991qq,Boyanovsky:1999jg,Durrer:2001cg,VilShe94,Hindmarsh:1994re,Hindmarsh:2011qj}. All cases, topological or not, local or global, are termed cosmic defects. 

Based on dimensional grounds and causality, it was argued in~\cite{Krauss:1991qu} that global phase transitions generate an approximately scale-invariant GW background. The amplitude was estimated with the quadrupole approximation for GWs, without any reference to the number $N$ of components of the  symmetry breaking field. In the context of the large $N$ limit of a global phase transition~\cite{Turok:1991qq,Boyanovsky:1999jg} and using a full treatment of the tensor metric perturbation representing GWs (i.e.~not resorting to the quadrupole approximation), it was demonstrated in~\cite{JonesSmith:2007ne,Fenu:2009qf} that indeed an exact scale-invariant background of GW is generated by the self-ordering process of the non-topological global textures arising after the phase transition. 

In this letter we generalize these results. We demonstrate that any scaling source in the radiation era produces a background of GWs with a scale-invariant energy density power spectrum. In the case of defects, we emphasise that the result is not related to their topology, the order of phase transition or the global/local nature of the symmetry-breaking process; it is just a consequence of scaling. As an example, using lattice simulations as an input, we calculate numerically the GW amplitude from a system of global O($N$) defects. We consider simulations with $N = 2$, $3$, $4$, $8$, $12$ and $20$, thus probing both topological and non-topological global defect scenarios. We provide evidence that the numerical results converge to the large $N$ calculation [Eq.~(\ref{eq:GW_largeN}) below] as $N^{-2}$, albeit with a large coefficient.  The GW power from global strings ($N=2$) is significantly above the trend, around 100 times greater than the large $N$ prediction.

On the numerical side, our present work is complementary to that recently presented in~\cite{Giblin:2011yh}, where numerical simulations are also performed and qualitative agreement with the large $N$ result is obtained. There, the authors conclude that global defects created after a second-order phase transition generate a scale-invariant GW background. Such a conclusion, however correct, is not fully supported by their numerical spectra: 
their GW spectra could be consistent with scale invariance, but there are large fluctuations. Our aim in this letter is precisely to demonstrate that scale-invariance is indeed exact for any scaling source, and that this is not related to the type of phase transition.

In the global O($N$) theory, although the field equations are non-linear, analytic calculations are possible in the limit of large $N$~\cite{Turok:1991qq,Boyanovsky:1999jg}, which show that the self-ordering dynamics of the non-topological defects exhibits scaling. It is also possible to calculate analytically the GW power spectrum \cite{Fenu:2009qf}, (see also \footnote{Note that actual dimensionless number characterizing the amplitude given in~\cite{Fenu:2009qf} is $511$. An improved numerical integration in~\cite{DaniPhD} gives 650.}), as 
\ben\label{eq:GW_largeN}
\Omgw(f) \equiv \frac{1}{\rho_{c}}\frac{d\rho_{_{\rm GW}}}{d\log f}(f) \simeq \frac{650}{N}\, \Omega_{\rm rad}\left(\frac{v}{{\Mpl}}\right)^{\!\!4},
\een
where  $\rho_c$ is the critical energy density today, and $\Mpl \approx 1.22\times10^{19}\,{\rm GeV}$ is the Planck mass.  The parameter $v$ is the vacuum expectation value (vev) of the scalar field and $\Omega_{\rm rad} \simeq 4\times10^{-5}$ is the radiation-to-critical energy density ratio today. The amplitude of this background does not depend on the frequency $f$, it is an exact scale-invariant background. There is no dependence either on the self-coupling $\lambda$ of the symmetry-breaking field. This is because the effective theory of the Goldstone modes, responsible for the creation of the GWs, 
is a non-linear $\si$-model, and the coupling disappears when the scalar field mode is integrated out.  

Important questions are raised, which we address in this letter.  
How does the scale-invariant GW spectrum come about? How does the GW spectrum look in the case of topological defects? 
How does the true GW signal from global non-topological textures approach the large $N$ result? The last two questions are 
particularly relevant in string-inspired models such as \cite{Dasgupta:2004dw,Burgess:2008ri}, which can have (approximate) global symmetries with low $N$.

In the following we shall assume that the total energy-momentum tensor $T_{\mu\nu}$ has contributions from ideal radiation, matter, and defects, so that we can split it as $T_{\mu\nu} = T_{\mu\nu}^{\rm rad} + T_{\mu\nu}^{\rm mat} + T_{\mu\nu}^{\rm def}$. Unlike radiation and matter, the energy-momentum tensor of defects is not a perfect fluid, and supports anisotropic stresses.

We assume that the defects create a small perturbation on a homogenous and isotropic cosmological background. The metric may  be written as a small departure from the Friedmann-Lema\^itre-Robertson-Walker (FLRW) form 
\begin{equation}
ds^2 = a^2(t)(\eta_{\mn}+h_{\mu\nu})dx^\mu dx^\nu,\end{equation} 
with $dx^0 = dt$ the conformal time and $a(t)$ the scale factor. GWs are represented by the transverse and traceless (TT) parts of the metric perturbations $h_{ij}^{{\TT}}$, satisfying $\partial_i h_{ij}^\TT = h_{ii}^\TT = 0$. Expanding the Einstein equations to first order in $h_{ij}^\TT$, we obtain
\begin{equation}\label{eq:GWeq}
\ddot{\bar{h}}_{ij}^{\TT} - \left(\nabla^2 + \frac{\ddot a}{a}\right){\bar{h}}_{ij}^{\TT} = \frac{16\pi a(t)}{\Mpl^2}\Pi_{ij}^\TT,
\end{equation} 
where $\bar h_{ij}^\TT \equiv a h_{ij}^\TT$, and $\Pi_{ij}^\TT$ is the TT part of $T_{ij}$.
The spectrum of energy density of a stochastic GW background in comoving momentum $k$ is given by
\begin{eqnarray}\label{eq:GWspectrum}
\frac{d\rho_{_{\rm GW}}}{d\log k}(k,t) &=& \frac{\Mpl^2k^3|{\dot h}_{k}(t)|^2}{64\pi^3a^2(t)}\,,
\end{eqnarray}
with $|{\dot h}_k(t)|^2$ the power spectrum of $\dot h_{ij}^\TT$. The solution in Fourier space to Eq.~(\ref{eq:GWeq}) is
\begin{equation}
 h_{ij}^{\TT}(k,t) = \frac{16\pi}{a(t)\,\Mpl^2}\int_{t_{i}}^t dt' a(t')\mathcal{G}(k,t,t')\Pi_{ij}^{\TT}(k,t')
\end{equation}
with $t_i$ the initial time of GW production, i.e.~$h_{ij}^\TT(t_i) = \dot{h}_{ij}^\TT(t_i) = 0$, and $\mathcal{G}(k,t,t')$ the retarded Green's function associated to the differential operator in the left hand side of Eq.~(\ref{eq:GWeq}). At sub-horizon scales ($kt,kt' \gg 1$), $\mathcal{G}(k,t,t') = k^{-1}\sin(k(t-t'))$. Averaging over a time $\delta t = {2\pi/k}$, the GW spectrum becomes
\begin{eqnarray}\label{eq:GWfromUETC}
\frac{d\rho_{GW}}{d\log k}(k,t) &=& \frac{2k^3}{\pi \Mpl^2}\frac{1}{a^4(t)}\int_{t_i}^t dt' \int _{t_i}^t dt''a(t') a(t'') \nn\\
&& \times\,\, \cos(k(t'-t''))\,\Pi^2(k,t',t''),
\end{eqnarray}
where ${\Pi}^2$ is the unequal time correlator (UETC) of $\Pi_{ij}^{\TT}$, 
\begin{equation}
\langle {\Pi}_{ij}^\TT\hspace*{-0.5mm}(\bk,t){{\Pi}_{ij}^{\TT}}^{\hspace*{-0.2mm}*}\hspace*{-1mm}(\bk',t')\rangle \!= \!(2\pi)^3{\Pi}^2(k,t,t')\delta_D(\bk-\bk')\,.
\end{equation}
The correlator $\Pi^2(k,t_1,t_2)$ can be obtained in general from field theory simulations. 
If the source is scaling, then the UETC can only depend on $k$ through the variables $x_1 = kt_1$ and $x_2=kt_2$. From dimensional analysis  
\begin{equation}
\Pi^2(k,t_1,t_2) = \frac{4v^4}{\sqrt{t_1t_2}}\,\uetcT(x_1,x_2),
\end{equation}
with  the factor 4 chosen so that $\uetcT$ agrees with the tensor UETC of Ref.~\cite{Bevis:2010gj}. Using this form of the correlator and the fact that in a radiation background the scale factor normalized to unity today can be written as $a(t) = \sqrt{\Omega_{\rm rad}}H_0\,t$, we obtain at sub-horizon scales $x \equiv kt \gg 1$, that the spectrum of GW becomes 
\begin{eqnarray}\label{eq:GWspectrum2}
\frac{d\rhogw}{d\log k}(x,t) = \Omega_{\rm rad}\frac{4}{\pi}\frac{\Mpl^2H_0^2}{a(t)^4}\,\left(\frac{v}{\Mpl}\right)^{\!\!4}\, \integral(x)\hspace*{0.7cm}\\
\label{eq:F_U}
\integral\!(x) \equiv 2\!\int^x\hspace*{-0.3cm} dx_1\!\!\int^x\hspace*{-0.3cm} dx_2 \sqrt{x_1x_2} \cos(x_1-x_2)\,\uetcT\!(x_1,x_2)
\end{eqnarray} 
At subhorizon scales $\uetcT(x_1,x_2)$ is peaked near $x_1 = x_2$, and decays along the diagonal as a power law (see e.g.~\cite{Durrer:2001cg}). It also decays away from the diagonal due to the lack of time coherence of the source \cite{Albrecht:1995bg}.
Hence the convergence of the integration is guaranteed for fast enough decays. That implies that $\integral(x)$ becomes more and more insensitive to its upper bound of integration, approaching asymptotically  a constant value for $x \gg 1$. In other words, $\integral_\infty = \integral(x\rightarrow \infty)$ is a  constant. As a consequence of this, the GW spectrum at subhorizon scales becomes scale-invariant.

For every type of defect there is indeed a function $\uetcT(x_1,x_2)$, and thus a well-determined value $\integral_\infty$, which characterizes the amplitude of the GW background. In particular, redshifting the amplitude today and using $3H_0^2\Mpl^2 = 8\pi\rho_{\rm c}$, we obtain
\begin{eqnarray}\label{eq:GWspectrumToday}
\Omgw(k) \equiv \frac{1}{\rho_{\rm c}}\left(\frac{d\rhogw}{d\log k}\right) = \frac{32}{3}\,\Omega_{\rm rad}\left(\frac{v}{\Mpl}\right)^{\hspace*{-1mm}4}{\hspace*{-0.5mm}}\integral_\infty\,.
\end{eqnarray}
That is, the background of \gws\ produced during the radiation era by the evolution of any network of defects in scaling regime, is exactly scale-invariant.  The amplitude is supressed by the fraction $\Omega_{\rm rad}$, and it depends on the vev as $(v/\Mpl)^4$, and on the shape of the UETC, which ultimately modulates the amplitude through $\integral_\infty$. We can identify the value of $\integral_\infty$ in the large $N$ analytical calculation~\cite{Fenu:2009qf} as $({32/3})\integral_\infty(N) \simeq {650/ N}$.

The method of calculating UETCs from lattice field theory simulations is well-documented \cite{Pen:1993nx,Pen:1997ae,Bevis:2006mj,Bevis:2010gj}. We consider a model with a global O($N$) symmetry, spontaneously broken to O($N-1$) in the ground state. We take a scalar field $\Phi$ with $N$ (real) components, $\Phi = (\varphi_1, ...,\varphi_N)^{\rm T}/\sqrt{2}$. The lagrangian of the model and the energy-momentum tensor are given by 
\begin{eqnarray}
\mathcal{L}(\Phi) = \partial_\mu\Phi^{\rm T}\partial_\mu \Phi - {\lambda}\left(\Phi^{\rm T}\Phi - v^2/2\right)^2\\
T_{\mu\nu}(\bx,t) = 2\partial_\mu\Phi^{\rm T}\partial_\nu\Phi - g_{\mu\nu}\, \mathcal{L}(\Phi)~~~~~
\end{eqnarray}
where $\Phi^{\rm T}\Phi \equiv {1/2}\sum_{m} \varphi_m^2$. After symmetry breaking the scalar is very close to its vacuum expectation value.

Taking the spatial Fourier transform of $T_{ij}$, the two tensor polarizations ($A = 1,2$) contributing to the GW source are defined as
\begin{eqnarray}
S^T_A(\bk,t)=\sqrt{\frac{t}{2}} \sum_{i,j} M^A_{ij}T_{ij}(\bk,t) ,
\end{eqnarray}
where the projectors $M^A_{ij}$ obey $\sum_{A} M^A_{ij}M^A_{lm} = \La_{ij,lm} \equiv P_{il}P_{jm} - {1/2} P_{ij}P_{lm}$, with $P_{ij} = \de_{ij} - \hat k_i \hat k_j$. Here $\La_{ij,lm}$ is the projector onto the TT part of $T_{ij}$. The UETC is 
\begin{equation}
\uetcT(x_1,x_2 ) =\frac{1}{2} \sum_A \vev{S^T_A(\bk,t_1 ) S^T_A(\bk,t_2 )^*}, 
\end{equation}
where the average is taken over a set of numerical simulations and a shell in Fourier space.
The UETC obeys the symmetry $\uetcT(x_1,x_2) = \uetcT(x_2,x_1)$.  

The algorithm for the numerical simulations solves the Klein-Gordon field equations obtained from the  lagrangian above, on a periodic Cartesian grid and using a 7-point stencil for the 3D-Laplacian and a leapfrog for the time evolution.  We keep $a^2\la$ constant to maintain a constant comoving scalar mass $m = \sqrt{\la}v$, with $\la=1$ and $v = 1$ \cite{Bevis:2006mj,Bevis:2010gj}. The grid size was $1024^3$, the spacing was $\De x = 0.5 $, and the timestep $\De t = 0.2\De x$. The fields are typically initialised at conformal time $t=1$ with independent random values constrained to lie on the $(N-1)$-sphere $\Phi^{\rm T}\Phi = v^2/2$, and with $\dot\Phi = 0$. The system is evolved initially with a period of variable diffusion or dissipation,  so that it  relaxes quickly to scaling. 
The UETCs are constructed by multiplying the Fourier transforms from 50 logarithmically  spaced times in the range $\tRef \le t_1 < \tEnd$, with the one taken at a  
reference time $t_2=\tRef$. For $N>2$, $(\tRef,\tEnd) = (64,232)$. Strings need more time to reach scaling, so 
for $N=2$, $(\tRef,\tEnd) = (150,300)$.
To obtain the power spectra we average over a shell of width $\De k = 2\pi/L$, where $L$ is the side length of the simulation volume. 

\begin{figure}[t]
\begin{center}
\includegraphics[width=0.5\textwidth]{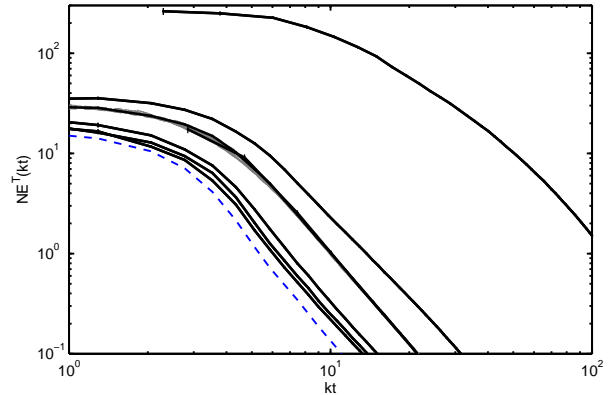}
\end{center}
\vspace*{-10mm}
\caption{ETCs from simulations, $N\etcT_{\rm num}(x)$ (solid black), evaluated at $t = 64$ ($t=150$ for $N=2$). 
From highest to lowest, we give $N$ = 2, 3, 4, 8, 12 and 20. Lowest of all is $N\etcT_{\rm th}(x)$ (dashed). For $N=4$ we also show the ETCs at $t=232$, and times in between in grey, to demonstrate the 
excellent scaling (for $N = 2$, due to later onset of scaling, 
we plot from $x \geq 1.84$). The error bars on the numerical curves give the 1$\si$ variation over all runs, and are barely visible.
} 
\label{fig1}
\end{figure}

To compare the dynamics in our lattice with the large $N$ analytical calculation, we computed the equal time correlator (ETC) $\etcT(x) = \uetcT(x,x)$ for each of $N$ = 2, 3, 4, 8, 12 and 20. The theoretical ETC in the large $N$ limit scales as $\etcT_{\rm th}(x)\propto1/N$. Thus, in Fig.~\ref{fig1} we plot $N\etcT_{\rm num}(x)$ at $\tRef$, obtained (averaging over 20 realizations) for each $N$, with $N\etcT_{\rm th}(x)$ for comparison. 

Computing the numerical to theoretical ratio of ETCs at the scale $x = \pi$ when one full wavelength enters the horizon, $\Upsilon_N \equiv \etcT_{\rm num}(\pi)/\etcT_{\rm th}(\pi)$, we find the values exhibited in Table \ref{tab:a}. The numerical ETC approaches the theoretical prediction as $N$ grows, as shown by the approach of $\Upsilon_N$ to unity as $N$ increases. For the case of cosmic strings ($N = 2$) the numerical ETC is a factor $\sim 40$ bigger than the theoretical one, signalling the breakdown of the large $N$ approximation. We do not expect the large $N$ approximation to apply for $N=2$: not only is $N$ small, but there are string defects which invalidate the mean-field analysis from the start.  The reason that the scaling density is so much larger than the large $N$ value is however unclear to us.

\begin{figure}[t]
\begin{center}
\includegraphics[width=0.5\textwidth]{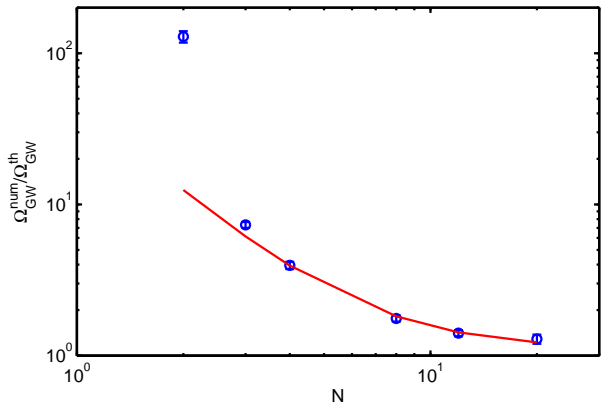}
\end{center}
\vspace*{-10mm}
\caption{Ratio of the numerical GW amplitude $\Omega_{\rm GW}^{\rm num}$ to the large $N$ analytical calculation $\Omega_{\rm GW}^{\rm th}$ (see Table \ref{tab:a}) and a fit to $1.1\! + 45/N^2\!$. The error bars give the 1$\si$ variation over all runs.}
\label{fig2}
\end{figure}

In Fig.~\ref{fig2} we compare the GW amplitude today $\Omega_{\rm GW}^{\rm th}$ given by the large $N$ theoretical calculation Eq.~(\ref{eq:GW_largeN}), versus the amplitude $\Omega_{\rm GW}^{\rm num}$ obtained from Eq.~(\ref{eq:GWspectrumToday}), with $\integral_{\infty}$ calculated from the numerical UETC for each scenario. We see that $\Omega_{\rm GW}^{\rm num}/\Omega_{\rm GW}^{\rm th}$ approaches unity as $N$ grows, showing that the numerical  amplitudes converge to the large $N$ analytical result. Today's amplitudes for each different $N$ considered are summarized in Table~\ref{tab:a}. If we parametrize the numerical amplitude for $N \geq 4$ as
\begin{equation}
\Omega_{\rm GW}^{\rm num} = \Omega_{\rm GW}^{\rm th}\left(a_0 + \frac{a_1}{N} + \frac{a_2}{N^2} + ...\right), 
\end{equation}
we find a good approximation with $a_0 \simeq 1.1$, $a_2 \simeq 45$ (see Fig.~\ref{fig2}) with negligible $a_1$. 
We believe that the $10\%$ deviation from unity of $a_0$ is due to a systematic uncertainty in our numerics, 
most likely a finite volume effect.
Thus, the numerical GW amplitude approaches the large $N$ result faster than naively expected, as $\propto 1/N^2$, albeit with a large coefficient.  The convergence reflects the behaviour of the overall scale of the UETCs, as measured by $\Upsilon_N$, although we see some $N$-dependence in the UETC width, which we shall report on in the future. Strings are well above this trend, by a factor of about 100.  We cannot be more precise at this stage, as there is a systematic uncertainty arising from the extrapolation of the UETC to $x=0$. We estimate this to be of order 50\%.

\begin{table}[!htp]
\begin{tabular}{|| c | c c c c c c||}
\hline
    {${N}$}
  & {$ 2 $}
  & {$ 3 $}
  & {$ 4 $}
  & {$ 8 $}
  & {$ 12 $} 
  & {$ 20 $}  \\
\hline
$\Upsilon_N$ & 36  & 4.5 & 3.1 & 1.7 & 1.4 & 1.3\\ 
\hline
$\Omega_{\rm GW}^{\rm num}/\Omega_{\rm GW}^{\rm th}$ & 130 & 7.3 & 3.9 & 1.8 & 1.4 & 1.3\\ 
\hline
\end{tabular}
\caption{Values of the numerical ETCs at $x=\pi$, and GW amplitudes today, normalized to the large $N$ calculation. The fluctuation in the amplitudes over the 20 realizations is less than 10\%, except for $N = 2$ where it is $ \sim 20\%$.
\label{tab:a}}
\end{table}

In this letter we have clarified the origin of the scale-invariance of the GW background calculated in~\cite{JonesSmith:2007ne,Fenu:2009qf} for the case of non-topological global textures. More importantly, we have generalized the result: a scale invariant background of GW is expected from any scaling cosmological source during the radiation era. In particular, global defects, independent of their topology, and cosmic gauged strings (local or semi-local), enter into a scaling regime, and produce a scale-invariant (i.e. frequency-independent) GW power spectrum according to Eq.~(\ref{eq:GWspectrumToday}), whose amplitude depends on the defect type.

We performed numerical simulations of the self-ordering dynamics of an O($N$) scalar field, showing that the GW power spectrum approaches the large $N$ prediction at a rate consistent with $N^{-2}$ (with a surprisingly big coefficient). For example, for $N=4$ the GW power spectrum is approximately four times larger than the large $N$ prediction. For strings, the factor is of order 100. 

We note that global strings  ($N\!\!=\!2$) decay by emission of massless \cite{Hiramatsu:2010yu} and massive scalar radiation, both from infinite strings and loops, at a rate proportional to $(v/\Mpl)^2$. Hence the GW emission, whose power is proportional to $(v/\Mpl)^4$, is not a significant source of energy loss. Global strings therefore do not behave like local strings in the Nambu-Goto approximation, which decay into GWs alone, via emission from sub-horizon size string loops. The amplitude of this background depends sensitively on the as-yet uncertain loop size distribution (see~\cite{Sanidas:2012ee,Sanidas:2012tf} for a recent well-referenced investigation).  We emphasise that the background we predict  arises from long strings and short-lived horizon-size loops, and has not been considered before. While subdominant for Nambu-Goto strings, it forms an irreducible minimum for strings decaying by particle emission.

It will be interesting to calculate the GW power spectrum from gauge cosmic strings, where numerical simulations show that the ETC decays much more slowly. The GWs can contribute appreciably to the relativistic energy density, with important implications for the cosmic microwave background power spectrum \cite{Lizarraga:2012mq}.

We are very grateful to Valerie Domcke for pointing out to us an important typo in the first version of the manuscript. DGF is supported by the Swiss National Science Foundation. MH acknowledges support from the Science and Technology Facilities Council (grant number ST/J000477/1). JU acknowledges support from the Basque Government (IT-559-10), the Spanish Ministry (FPA2009-10612) and the  Consolider-Ingenio  Programme CPAN (CSD2007-00042). Numerical calculations were performed using the UK National Cosmology Supercomputer (supported by SGI/Intel, HEFCE, and STFC).

\bibliography{GWO}

\end{document}